\documentclass[aip,jap,reprint,amsmath,amssymb,longbibliography]{revtex4-2}
\usepackage[T1]{fontenc}
\usepackage{lmodern}
\usepackage[utf8]{inputenc}
\usepackage{graphicx,bm,booktabs}
\usepackage{xcolor}
\usepackage[colorlinks=true,linkcolor=blue!45!black,citecolor=blue!45!black,urlcolor=blue!45!black]{hyperref}
\newcommand{\sgn}{\operatorname{sgn}}
\newcommand{\diag}{\operatorname{diag}}
\newcommand{\ii}{\mathrm{i}}
\newcommand{\dd}{\mathrm{d}}
\newcommand{\Id}{\mathbb{I}}
\hypersetup{pdftitle={Sublattice-selective control of spin reversal in metasurface-coupled Kagome interfaces}}
\begin{document}
\title{Sublattice-selective control of spin reversal in metasurface-coupled Kagome interfaces}
\author{Ximo Wang}
\author{Qiwei Han}
\author{Zhenqi Bai}
\author{Ruyue Guo}
\author{Min Feng}
\author{Yichi Zhang}
\email{zhangyichi@sxu.edu.cn}
\affiliation{College of Physics and Electronic Engineering, Shanxi University, 030006 Taiyuan, People’s Republic of China}
\affiliation{Collaborative Innovation Center of Extreme Optics, Shanxi University, Taiyuan, Shanxi 030006, People’s Republic of China}
\date{15 September 2026}
\begin{abstract}
Coherent manipulation of photonic interface states requires a control field that matches their internal mode structure. We study this requirement in a two-component Kagome lattice motivated by photon-mediated exchange near a nonlinear nonlocal metasurface. The Dirac spinors show why uniform Raman control cannot couple opposite-spin, same-valley interface modes at leading order, even when their spatial envelopes coincide. A $1:1:-2$ sublattice pattern removes this cancellation and maximizes the projected coupling at fixed root-mean-square amplitude within the diagonal-control class. We test this control scheme through full-zone topology and complete lattice propagation. For a smooth, gapped interface, a $720$-dimensional calculation gives target-mode fidelity $0.999937$ with leakage $6.27\times10^{-5}$ at RMS drive $0.04t$. Trace-preserving dynamics gives the separate survival condition needed for successful conversion. An exploratory three-dimensional lithium-niobate supercell reproduces the complex addressing pattern with $0.373\%$ relative error and provides nonlocal exchange, decay and electro-optic frequency-conversion matrices. The mode-resolved control principle thus gives quantitative electromagnetic design targets; the full spin-dependent device realization still requires further calibration.
\end{abstract}
\maketitle

\section{Introduction}
Photonic interfaces provide a route to guide optical excitations through band engineering. Chiral boundary propagation was established by theoretical proposals and microwave experiments,\cite{HaldaneRaghu2008,RaghuHaldane2008,Wang2009} followed by synthetic gauge fields and spin-like channels in ring resonators, modulated waveguides and dielectric structures.\cite{Hafezi2011,Hafezi2013,Rechtsman2013,WuHu2015} This work connects topological band geometry to optical transport.\cite{Lu2014,Ozawa2019} Active manipulation also requires a field that couples the particular interface modes to be addressed. Valley-selective excitation and routing in bianisotropic, dielectric and surface-plasmon crystals show how the source and internal mode structure jointly determine the observed response.\cite{CNTopoDong2017,CNTopoChen2017,CNTopoWu2017,CNTopoHe2019} Counterpropagating channels alone do not specify how to coherently convert one channel into the other.

The Kagome lattice makes this control problem explicit. Its three-site basis combines Dirac crossings with a flat band whose localized states arise from destructive interference.\cite{Bergman2008,Leykam2018,Zong2016} Complex hopping supports topological Kagome bands and photonic Hall responses, while related models connect band flatness with nontrivial geometry.\cite{Ohgushi2000,Guo2009,Petrescu2012,Tang2011,Sun2011} Experiments on flat-band line and loop states,\cite{CNTopoXia2018,CNTopoMa2020} sublattice-dependent protection,\cite{CNTopoWang2023} and higher-order photonic corner states\cite{CNTopoXie2019,CNTopoChen2019,CNTopoZhong2024} also show why amplitudes within a unit cell matter. For a spin-dependent Dirac interface, this internal structure can suppress a transition even when the two modes occupy the same spatial region. The overlap of their intensity envelopes therefore does not determine the required control operator.

Photon-mediated emitter arrays offer a microscopic setting for these interfaces. Their coherent exchange and collective radiation follow from the same electromagnetic Green tensor, so the photonic environment can shape their interactions.\cite{Douglas2015,GonzalezTudela2015,Perczel2017,Perczel2020} Nonlinear metasurfaces offer a possible route to optical control of this response: dielectric resonators enhance frequency conversion,\cite{Liu2016SHG,Vabishchevich2018,Fedotova2020} and nonlocal resonances extend field manipulation across multiple unit cells.\cite{Overvig2020,Malek2022} These capabilities motivate a metasurface-coupled architecture with jointly designed relative phases and amplitudes. Converting a desired sublattice pattern into incident optical fields then requires retaining collective scattering.

Raman control offers established ways to couple internal states and engineer momentum-dependent interactions. Spin-orbit-coupled Fermi gases, two-electron atoms and optical Raman lattices demonstrate control over the amplitude, phase and spatial structure of these couplings.\cite{CNRamanWang2012,CNRamanHuang2016,CNRamanSong2016,CNRamanWu2016,CNRamanSun2018Robust} Raman-assisted flux patterns and three-dimensional Weyl-band implementations extend this control to more complex operators.\cite{CNRamanLiu2016,CNRamanWang2021} In the present array, the relevant matrix element must be evaluated between the actual Kagome interface spinors. A uniform spin rotation need not drive a transition within the selected interface subspace, and a large total spin change need not mean transfer to the intended valley.

\begin{figure*}[t]
\centering\includegraphics[width=\textwidth]{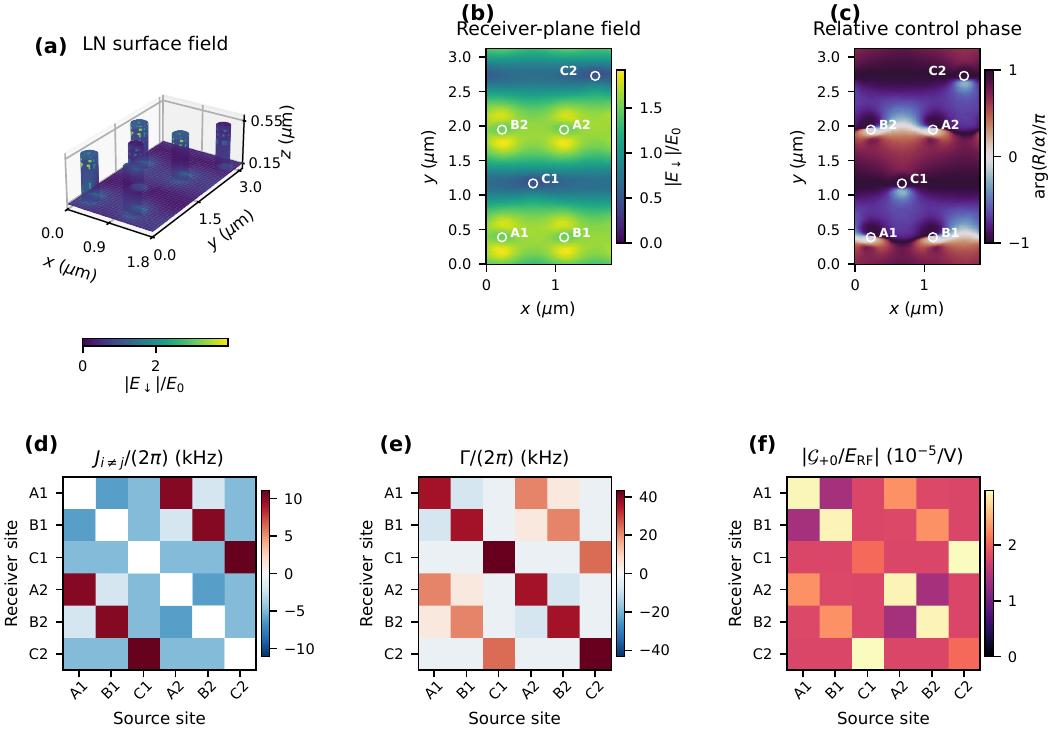}
\caption{Exploratory COMSOL supercell and extracted response. (a) Three-dimensional electric-field magnitude on the LN film and six pillar surfaces, evaluated just inside the dielectric. (b) Field magnitude and (c) Raman-product phase at the receiver plane, $100$ nm above the pillars. White circles locate the six finite-volume receivers. Phase is relative to the fitted common complex scale $\alpha$; $\pm1$ denotes a $\pi$ offset. Fields use the optimized down-control solution with $E_0=1$ V/m and 40-layer PMLs. (d) Regularized coherent exchange and (e) collective decay, from the full matrix-adjoint decomposition; the real diagonal exchange shift is omitted from the plot. Exchange and decay use an illustrative 1-Debye transition dipole. (f) Upper-sideband Green response per prescribed RF field. Matrices are $k_\parallel=0$, lattice-summed, finite-volume $zz$ responses, with 20-layer PMLs and a 40-layer source-column check. All panels derive from FEM data; material constants, dimensions and RF amplitude are exploratory assumptions.}
\label{fig:comsol}
\end{figure*}

We derive and test a sublattice-selective solution to this problem. For an explicit two-component Kagome Hamiltonian, we obtain the uniform-control selection rule and a fixed-RMS bound saturated by the complex pattern $1:1:-2$. Complete strip calculations retain both folded valleys and quantify target-mode transfer, leakage and momentum acceptance. We then use a passive open-system treatment to separate conditional conversion from excitation survival. A three-dimensional COMSOL supercell tests the required complex optical addressing and supplies exchange, decay and second-order electro-optic response data (Fig.~\ref{fig:comsol}). The effective-lattice analysis sets the control target, and the electromagnetic model tests a specified exploratory design.

\section{Effective model and control selection rule}
\subsection{Electromagnetic input and excitation encoding}
We define two retained excitation flavors $\sigma=\uparrow,\downarrow$ above a separate vacuum, either as metastable excitations with eliminated auxiliaries or as weak-probe amplitudes. A ground-state Raman rotation of fixed two-level atoms does not automatically preserve this encoding. In angular-frequency units, passive conditional dynamics is generated by
\begin{equation}
 H_{\rm cond}=H_{\rm coh}-\ii K/2,\quad H_{\rm coh}=H_{\rm coh}^\dagger,\quad K\succeq0.
 \label{eq:conditional}
\end{equation}
Unconditional probabilities additionally require jump terms.\cite{Lindblad1976,Dalibard1992}

Using the $e^{-\ii\omega t}$ convention, the outgoing Green tensor obeys $[\nabla\times\nabla\times-(\omega^2/c^2)\epsilon_b]G_b=\Id\delta$, with relative permittivity $\epsilon_b$. Stationary Markovian exchange then gives\cite{Lehmberg1970,Dung1998,Dung2002,AsenjoGarcia2017Green,Manzoni2018}
\begin{align}
 \Sigma_{i\sigma,j\sigma'}&=-\frac{\mu_0\omega_0^2}{\hbar}\mathbf d_{i\sigma}^*\cdot G_{ij}(\omega_0)\cdot\mathbf d_{j\sigma'},\nonumber\\
 J&=(\Sigma+\Sigma^\dagger)/2,\quad \Gamma=\ii(\Sigma-\Sigma^\dagger).
 \label{eq:sigma}
\end{align}
The local real ultraviolet shift is renormalized. The adjoint acts on the full site/flavor matrix; entrywise real and imaginary parts cannot replace it.

For frequency channels $\omega_n=\omega+n\omega_p$, the first Born conversion term is
\begin{equation}
 \mathcal G^{(1)}_{nm}=G_b(\omega_n)\frac{\omega_n^2}{c^2}\delta\epsilon_{nm}G_b(\omega_m),\quad n\ne m.
 \label{eq:floquet}
\end{equation}
Spatial integrations are implicit. The relative-permittivity kernel contains the pump and susceptibility contraction, with no extra $\epsilon_0$; a dipole source contributes $\mu_0\omega_m^2$. Carrier-preserving exchange ordinarily requires conversion out and back, a static bias, or an explicitly retained resonant channel. Thus $\chi^{(2)}$ alone does not imply a universal linear relation between pump amplitude and a carrier-frequency Dirac mass. Eliminating an auxiliary block gives $H_{s,\rm eff}(E)=H_s-V^\dagger(H_a-E)^{-1}V$ and requires a bounded, spectrally separated resolvent and corresponding effective jumps.\cite{ReiterSorensen2012,Brion2007} The microscopic conventions are given in Supplementary Sec.~S1.

\subsection{Specified Kagome lattice}
We specify a target finite-range operator without identifying it with Fig.~\ref{fig:comsol}. Let $\mathbf a_1=a(1,0)$, $\mathbf a_2=a(1/2,\sqrt3/2)$ and $(\mathbf r_A,\mathbf r_B,\mathbf r_C)=(0,\mathbf a_1/2,\mathbf a_2/2)$. For the upper entries $AB,AC,BC$,
\begin{align}
 [h_\sigma(\mathbf k)]_{\alpha\beta}={}&-2t\cos(\mathbf k\cdot\mathbf u_{\alpha\beta})\nonumber\\
 &-2\ii\sigma\lambda\nu_{\alpha\beta}\cos(\mathbf k\cdot\mathbf w_{\alpha\beta}),
 \label{eq:hopping}
\end{align}
with $t>0$, real $\lambda$, zero diagonal and Hermitian-conjugate lower entries. We use $\mathbf u=(\mathbf a_1/2,\mathbf a_2/2,(\mathbf a_2-\mathbf a_1)/2)$, $\mathbf w=(\mathbf a_2-\mathbf a_1/2,\mathbf a_1-\mathbf a_2/2,(\mathbf a_1+\mathbf a_2)/2)$, and $\nu=(1,-1,1)$. In spin-block order, the coherent Hamiltonian is
\begin{equation}
 H_6=\begin{pmatrix}h_++\Delta\Id_3/2&R/2\\R^\dagger/2&h_--\Delta\Id_3/2\end{pmatrix}.
 \label{eq:h6}
\end{equation}
The factor $1/2$ separates the Rabi amplitude from the Hamiltonian matrix element. For $\lambda=0$, the energies are $2t$ and $-t\pm t\sqrt{3+2S(\mathbf k)}$, where $S=\cos(\mathbf k\cdot\mathbf a_1)+\cos(\mathbf k\cdot\mathbf a_2)+\cos[\mathbf k\cdot(\mathbf a_2-\mathbf a_1)]$.

At $\mathbf K_\eta=(\eta4\pi/3a,0)$, the orthonormal basis $U=((1,-1,0)^T/\sqrt2,(1,1,2)^T/\sqrt6)$ gives
\begin{equation}
 h^D_{\sigma\eta}=-t\Id_2+\eta v(-q_xs_z+q_ys_x)+\mu_\sigma s_y+O(q^2),
 \label{eq:dirac}
\end{equation}
where $v=\sqrt3at/2$ and $\mu_\sigma=2\sqrt3\sigma\lambda$. A diagonal potential instead projects as
\begin{equation}
 U^\dagger\diag(d_A,d_B,d_C)U=d_0\Id_2+d_xs_x+d_zs_z,
 \label{eq:diagonal}
\end{equation}
with $d_0=(d_A+d_B+d_C)/3$, $d_x=(d_A-d_B)/(2\sqrt3)$ and $d_z=(d_A+d_B-2d_C)/6$. It generates no first-order $s_y$ mass. Uniform spin detuning shifts spin sectors without changing this sublattice mass.

\begin{figure*}[t]
\centering\includegraphics[width=\textwidth]{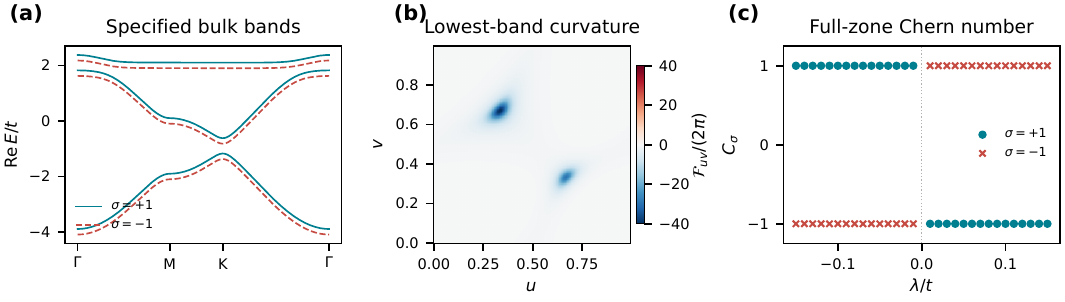}
\caption{Effective-model bulk verification. (a) Real bands for $\lambda/t=0.08$, $\Delta/t=0.20$; common loss gives $\operatorname{Im}E/t=-0.06$ for every band. (b) Lowest-band Berry-curvature density on a $151\times151$ periodic-gauge grid, $\mathbf k=u\mathbf b_1+v\mathbf b_2$, for spin $+$. (c) Full-zone lower-band Chern numbers for 30 nonzero signed couplings. No individual-band invariant is assigned at $\lambda=0$. These are lattice-model calculations, independent of the FEM supercell.}
\label{fig:bulk}
\end{figure*}

For $A=\ii\langle u|\nabla_{\mathbf k}u\rangle$, the small-mass full-zone continuation gives the lower-band invariant $C_\sigma=-\sigma\sgn\lambda$.\cite{Berry1984,Xiao2010,Thouless1982,Haldane1988} FHS links\cite{Fukui2005} yield spin-$+$ Chern numbers $(-1,2,-1)$ at $\lambda/t=0.08$, reversed for spin $-$. The lower direct gap is $0.5542562584t$. Independent Kubo-curvature integration converges to the same sign and integer (Supplementary Sec.~S5). Figure~\ref{fig:bulk} also shows $R=0$, $\Delta/t=0.20$, $K=0.12t\Id_6$, for which all six imaginary energies are $-0.06t$. Scalar loss preserves the eigenvectors and creates no exceptional point. The opposite spin Chern numbers describe conserved blocks; a quantum-spin-Hall interpretation also requires a protecting time-reversal representation.\cite{KaneMele2005QSH,KaneMele2005Z2} Arbitrary complex hopping needs its own gap and boundary analysis.\cite{Bergholtz2021}

\subsection{Sublattice projection and the fixed-RMS optimum}
For a sign wall $\lambda(x)=\lambda_0\sgn x$, the modes have envelope $\varphi_\sigma=\sqrt{\kappa_\sigma}e^{-\kappa_\sigma|x|}$, $\kappa_\sigma=|\mu_{\sigma,0}|/v$, and spinors satisfying $s_x|\xi_{\sigma\eta}\rangle=-\eta\sgn(\mu_{\sigma,0})|\xi_{\sigma\eta}\rangle$.\cite{JackiwRebbi1976} Their energies relative to $-t$ are $-\sgn(\mu_{\sigma,0})vq_y$. The Chern change is two, giving two valley channels per spin. A two-state reduction treats one selected valley; smooth controls and residual intervalley dynamics require separate checks.

The projected Rabi amplitude is
\begin{equation}
 \Omega_e=\int\dd x\,\varphi_\uparrow^*\varphi_\downarrow\langle\xi_\uparrow|U^\dagger R(x)U|\xi_\downarrow\rangle.
 \label{eq:overlap}
\end{equation}
Opposite-spin same-valley states have opposite $s_x$ eigenvalues. A uniform operator $R(x)=r(x)\Id_3$ therefore gives $\Omega_e=0$ at leading Dirac order, even when the envelopes overlap. Lattice corrections can give a finite residual, which we calculate below.

For constant diagonal $R=\diag(r_A,r_B,r_C)$ and sign-wall envelopes,
\begin{align}
 \Omega_e&=F\frac{r_A+r_B-2r_C}{6},\quad
 F=\frac{2\sqrt{\kappa_\uparrow\kappa_\downarrow}}{\kappa_\uparrow+\kappa_\downarrow}\le1,\nonumber\\
 |\Omega_e|&\le F r_{\rm rms}/\sqrt2,\quad
 r_{\rm rms}^2=(|r_A|^2+|r_B|^2+|r_C|^2)/3.
 \label{eq:bound}
\end{align}
The pattern $R=(r_{\rm rms}/\sqrt2)\diag(1,1,-2)$ saturates this bound; the minus sign denotes a $\pi$ phase, not a negative intensity. The optimum applies only to diagonal sublattice controls with a common transverse profile. For a finite half-window $L$, multiply the envelope factor by $1-e^{-(\kappa_\uparrow+\kappa_\downarrow)L}$. For other profiles, use Eq.~\eqref{eq:overlap}.

\section{Full-lattice and open-system tests}
\begin{figure*}[t]
\centering\includegraphics[width=\textwidth]{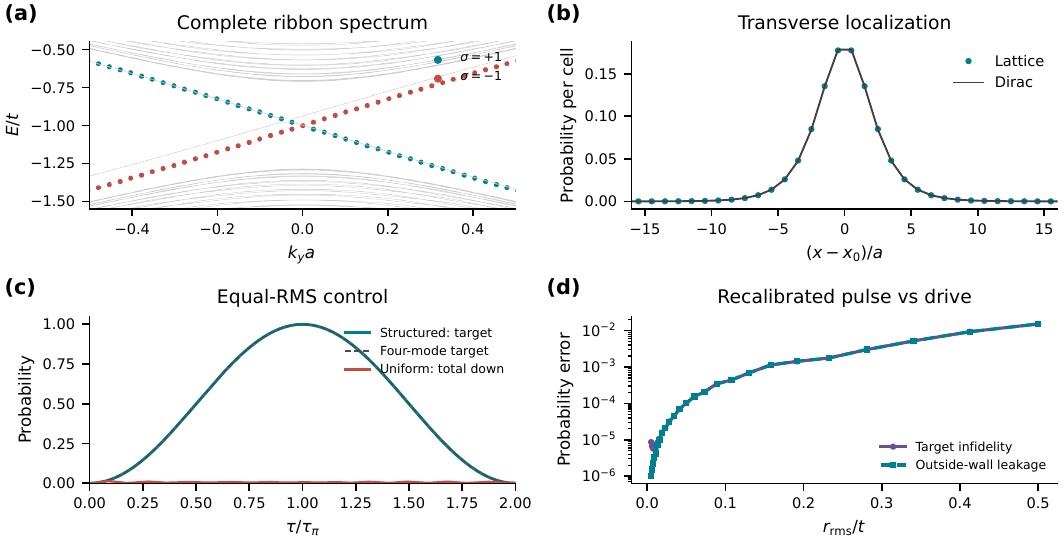}
\caption{Complete transverse-lattice tests at $\lambda_0/t=0.08$, $w/a=2$, $N_x=60$. (a) Ribbon spectrum; colored points have more than $80\%$ central probability, while gray curves include bulk and outer-boundary states. (b) Pair-averaged cell probability and normalized Dirac prediction. (c) Structured-drive target probability, four-mode prediction and uniform-drive total down-spin population at equal RMS $0.04t$. (d) Target infidelity and outside-wall leakage versus drive, with the projected pulse recalibrated at every point. These are conserved-momentum sectors, not a localized two-dimensional wavepacket.}
\label{fig:lattice}
\end{figure*}

We use a strip open in $x$ and periodic along $(0,\sqrt3a)$, with six sites per rectangular cell. We evaluate $\lambda(x)=0.08t\tanh[(x-x_0)/(2a)]$ at each next-nearest-neighbor bond midpoint. At $N_x=60$, two principal wall states per spin span a four-dimensional subspace; both valleys fold onto $k_y=0$. The analytic velocity operator gives spin-up group velocities of $-0.8660273ta$ and $-0.8660252ta$, consistent with $-\sqrt3ta/2$. Their density agrees with the smooth-wall envelope [Fig.~\ref{fig:lattice}(a,b)].

For structured and uniform controls at equal RMS, the complete coupling between opposite-spin wall subspaces has singular values $(0.70709588,0.70709538)$ and $(0.00192261,0.00188168)$. We obtain valley-labelled states by projecting continuum trials onto the exact wall subspaces and polar-orthonormalizing. The trial weight is $0.99863445$; we retain the valley-state energy residual $3.11\times10^{-6}t$ in four-mode dynamics. The chosen pattern reaches $0.9999964$ of the finite-lattice diagonal-control optimum.

At $r_{\rm rms}/t=0.04$, the projected $\pi$ pulse has $t\tau_\pi=111.073825$. Full $720\times720$ propagation yields down-spin population $0.99996033$, target fidelity $0.99993719$, outside-wall-subspace leakage $6.2693\times10^{-5}$, and opposite-valley probability $1.19\times10^{-7}$. With the same RMS and duration, uniform control yields down-spin population $0.00685249$ and target fidelity $2.1477\times10^{-5}$ [Fig.~\ref{fig:lattice}(c)]. Survival is unity in these coherent calculations. This comparison uses a fixed protocol and does not optimize arbitrarily long uniform pulses.

Sparse exponentiation, DOP853 integration and an independent integer-coordinate bond construction reproduce the baseline dynamics. Increasing $N_x$ from 60 to 84 changes target fidelity by less than $1.5\times10^{-6}$. Leakage grows with drive strength [Fig.~\ref{fig:lattice}(d)]. For a sharper wall, $(\lambda_0/t,w/a)=(0.12,0.75)$, spin population stays at $0.99995$ but target fidelity drops to $0.91556$, as the other valley acquires $0.08438$ population. Free evolution alone gives $0.08436$, so intrinsic valley precession dominates. Finite width accounts for the low-mass corner of the scan. Supplementary Secs.~S6--S7 give detailed convergence and control-error data. At the baseline fixed pulse, the central $90\%$ target window is $|k_ya|\le0.0052660$; Sec.~S8 checks this protocol-specific bandwidth.

\subsection{Conditional conversion and survival}
Non-Hermitian spectra depend on the full complex operator and its gap structure.\cite{Shen2018,Gong2018,Kawabata2019} Write a constant projected matrix as $H_e=\zeta\Id_2+A$, where
\begin{equation}
 A=\begin{pmatrix}z&b\\c&-z\end{pmatrix},\quad Q^2=z^2+bc,\quad A^2=Q^2\Id_2.
 \label{eq:edgeH}
\end{equation}
Then $e^{-\ii H_e\tau}=e^{-\ii\zeta\tau}[\cos(Q\tau)\Id_2-\ii\sin(Q\tau)A/Q]$, with $\sin(Q\tau)/Q\to\tau$ at $Q=0$. An exceptional point requires $Q^2=0$ and $A\ne0$.\cite{Miri2019,ElGanainy2018} Starting with spin up, set $C=\cos(Q\tau)$ and $S=\sin(Q\tau)/Q$; then
\begin{align}
 N&=e^{2\operatorname{Im}\zeta\tau}(|C-\ii zS|^2+|cS|^2),\nonumber\\
 F_{\downarrow|s}&=\frac{|cS|^2}{|C-\ii zS|^2+|cS|^2},\quad P_\downarrow=NF_{\downarrow|s}.
 \label{eq:success}
\end{align}
The last identity assumes loss into an absorbing vacuum; recycling needs the full master equation. The coefficient formulas require an orthonormal, spin-pure basis. Complete conditional transfer needs $C-\ii zS=0$ at a real positive time with nonzero output. For $b,c\ne0$, survival at that time is $e^{2\operatorname{Im}\zeta\tau}|c/b|$ (Supplementary Sec.~S4).

For $z=\delta+\ii\gamma$, $b=c^*=g$ real and $\zeta=\epsilon_e-\ii\kappa_0/2$, independent jumps to vacuum have rates $\kappa_0\mp2\gamma$. Passivity requires $\kappa_0\ge2|\gamma|$. On compensated resonance with common loss, $g=|\Omega_e|/2$, $\tau_\pi=\pi/|\Omega_e|$ and $P_\downarrow(\tau_\pi)=e^{-\kappa_0\tau_\pi}$. A fixed-phase shaped pulse satisfies $\int|\Omega_e(\tau)|\dd\tau=\pi$ only at zero detuning and with a stationary projected basis.

\begin{figure}[t]
\centering\includegraphics[width=\columnwidth]{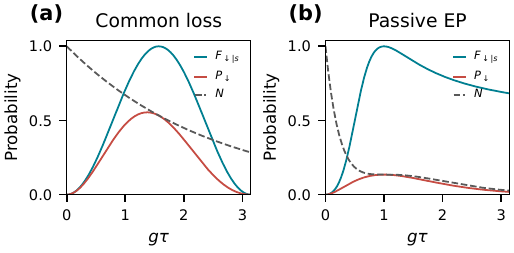}
\caption{Trace-preserving three-state dynamics: conditional conversion $F_{\downarrow|s}$, unconditional down-state probability $P_\downarrow$ and survival $N$. (a) $\gamma=0$, $\kappa_0/g=0.4$, $\delta=0$; at $g\tau=\pi/2$, $P_\downarrow=0.533488$. (b) Passive EP, $\gamma/g=-1$, $\kappa_0/g=2$; complete conditional conversion at $g\tau=1$ has success $e^{-2}=0.135335$.}
\label{fig:loss}
\end{figure}

An independent three-state Lindblad solver verifies Eq.~\eqref{eq:success} and its vacuum completion (Fig.~\ref{fig:loss}). The central $90\%$ conditional detuning window is $|\delta|/g\le0.322593$; $90\%$ unconditional success on resonance requires $\kappa_0/g\le0.067075$. The density matrix matches the no-jump completion and an independent ODE within $2.0\times10^{-12}$. Faster conditional transfer at a passive EP need not raise the success probability. Spin reversal is not locked to a bulk transition: as $\lambda\to0$, the localization length diverges and the isolated-edge approximation fails.

\section{Exploratory electromagnetic realization}
We use COMSOL~6.4 for an assumed two-cell LN supercell, $L_x=a=1.8\,\mu$m, $L_y=\sqrt3a$. Film and pillar heights are 150 and 400 nm; A/B and C radii are 160 and 140 nm. The assumed lossless indices are $(n_o,n_e,n_s)=(2.21,2.14,1.444)$ for $z$-axis LN and silica. The 80-nm source/receiver cubes sit 100 nm above the pillars. Floquet side boundaries at $k_\parallel=0$ and adapted vertical PMLs retain periodic multiple scattering. The dimensions and material constants serve as exploratory inputs (Supplementary Sec.~S11).

The up control at 1550 nm uses order $(0,+G_y)$. We jointly optimize three coherent down-control orders, $(0,+G_y)$, $(0,-G_y)$ and $(+G_x,0)$, at a 6.8-GHz offset with fixed incident-power normalization. We minimize the six-site complex residual to $\alpha(1,1,-2,1,1,-2)$, fitting a common scale $\alpha$. All sublattices are therefore coupled through the same solved field. Conjugating COMSOL's $e^{+\ii\omega t}$ fields to the paper convention gives the local Raman-product shape $R_j\propto\langle E_{\uparrow,z}\rangle_j\langle E_{\downarrow,z}\rangle_j^*$ in COMSOL variables, before a common atomic prefactor.

The optimized solution has $0.373\%$ relative complex-pattern error, $0.548\%$ maximum amplitude error and $0.224^\circ$ maximum phase error. Direct coherent and superposed fields agree to $6.7\times10^{-11}$. Refining the physical mesh with the original PML and refining the adapted PML from 20 to 40 layers change the control product by $0.940\%$ and $1.058\%$, respectively. These checks are separate; the first does not establish mesh convergence of the final 40-layer setup. The analytic vacuum-field check has $0.455\%$ error. Six dipole solves give a lattice-summed $6\times6$ $G_{zz}$ matrix with reciprocity residual $1.11\times10^{-13}$ and positive decay eigenvalues. These finite checks do not establish a full vector-spin Green tensor or determine the finite-range $t,\lambda$ model.

Second-order postprocessing uses a prescribed $z$-directed 6.8-GHz RF field with peak amplitude $E_{\rm RF}=10^5$ V/m. Nominal $r_{13}=8.6$ pm/V and $r_{33}=30.8$ pm/V specify the assumed small-signal EO model,
\begin{align}
 \delta\epsilon_{xx}^{+}=\delta\epsilon_{yy}^{+}&=-n_o^4r_{13}E_{\rm RF}/2,\nonumber\\
 \delta\epsilon_{zz}^{+}&=-n_e^4r_{33}E_{\rm RF}/2.
 \label{eq:eo}
\end{align}
The polarizations $P_\pm=\epsilon_0\delta\epsilon^\pm E_0$ generate sidebands. A return solve uses $P_{\rm ret}=\epsilon_0(\delta\epsilon^-E_++\delta\epsilon^+E_-)$. Doubling the RF field doubles the first sideband and quadruples the return field; field residuals stay below $3.2\times10^{-11}$. The extracted response norm is $\|\mathcal G_{+0}/E_{\rm RF}\|_F=1.2323\times10^{-4}$ V$^{-1}$. The full kernel uses 20-layer PMLs; checking one column with 40 layers changes it by $0.212\%$. This is small-signal EO frequency conversion, not 775-nm second-harmonic generation or an inversion of intrinsic $\chi^{(2)}$. We evaluate it separately from the unpumped Raman-pattern optimization.

The calculation does not establish RF electrode calibration, the selected atomic encoding, spin-dependent flux or the loss budget. Material dispersion, source-size convergence, full Bloch sampling and pump-dressed addressing need further tests before the FEM structure can be identified with the effective lattice. Nonlocal multiple scattering alone does not supply nonreciprocity or a topological mass. Supplementary Sec.~S11 gives the geometry, coefficients, extraction conventions and convergence data.

\section{Discussion}
The selection rule shows why the control must match the internal mode structure. A spatially uniform Raman field acts as the identity in sublattice space, giving a vanishing matrix element between the opposite interface spinors. The $1:1:-2$ pattern supplies the missing component, with a $\pi$ phase on C and no common sublattice contribution. The fixed-RMS optimum applies to diagonal controls with a common transverse profile; it does not establish optimality over arbitrary spatial operators or pulse sequences. Agreement with the exact smooth-wall target shows that the reduced result remains useful at the tested lattice parameters. The sharp-wall example shows why target fidelity and total down-spin population must be reported separately. Spin-texture and quench measurements can independently test bulk topology,\cite{CNRamanSong2018,CNRamanZhang2018,CNRamanSun2018Quench} whereas coherent interface conversion needs its own mode-resolved diagnostic.

Static spin reversal and pulse-driven transfer are different observables. For a nonzero right eigenvector $(b,Q-z)^T$ of the traceless two-state matrix, the normalized polarization is
\begin{equation}
 P_z^R=\frac{|b|^2-|Q-z|^2}{|b|^2+|Q-z|^2}.
 \label{eq:static}
\end{equation}
The polarization balances when $|Q-z|=|b|$. For coherent Hermitian coupling, $z=\delta\in\mathbb R$, $c=b^*$ and the positive real branch of $Q$, this gives $P_z^R=\delta/\sqrt{\delta^2+|b|^2}$. Preparing or following the eigenstate still requires a dynamical protocol. If a control changes both detuning and hopping, the polarization balance need not coincide with the bulk gap closing. The pulse protocol should keep the interface gapped, since the localization length diverges as the Dirac mass vanishes.

The FEM calculation turns this operator requirement into a collective optical addressing problem. Optimizing all six receiver amplitudes in one supercell includes the mutual response of A, B and C; phases of isolated pillars would not test this response. The target residual is smaller than the approximately $1\%$ changes in the mesh and PML checks, so the fitting precision must be read alongside those finite convergence estimates. The exported Green matrices retain lattice sums and finite receiver volumes. They fit within the framework used for photonic-crystal atom--light interfaces,\cite{Goban2014,Goban2015,Hood2016,Chang2018} cooperative arrays,\cite{Shahmoon2017,Rui2020} and subradiant storage and transport.\cite{AsenjoGarcia2017Subradiance,Needham2019} Relating them to the prescribed finite-range spinful lattice also requires polarization-resolved sampling, a controlled hopping truncation and a physical excitation encoding. Dirac reservoirs and control-assisted spin models offer relevant precedents,\cite{Perczel2020Dirac,Hung2016} while weak-probe descriptions and structured-reservoir memory impose further constraints.\cite{Fleischhauer2005,GonzalezTudela2017}

Nonlinear enhancement and the required exchange operator call for separate design choices. High-$Q$ and bound-state-in-the-continuum resonances offer routes to stronger optical conversion,\cite{Koshelev2018BIC,Koshelev2019Nonlinear,Carletti2018} including silicon and LN implementations or proposals.\cite{CNMetaLiu2019,CNMetaMa2021,CNMetaZhang2022,CNMetaLiu2023,CNMetaWang2024} Nonlocal wavefront control and coupled-mode descriptions\cite{Overvig2022,Overvig2024} complement angle-tunable harmonic conversion,\cite{CNMetaJiang2024SHG,CNMetaJiang2024THG} thermal nonlinear response and nonlinear nonlocal metasurfaces.\cite{Cotrufo2024,Cotrufo2026} Nonlinear phase and polarization control also demonstrate the role of complex fields.\cite{CNMetaLi2015,CNMetaLi2017,CNMetaMcDonnell2021} In our calculation, the assumed EO tensor gives frequency conversion at first Born order and a carrier return at second order. Doubling the RF field gives the expected factors of two and four, verifying the implemented scaling. These checks neither extract an intrinsic susceptibility from measured data nor establish a linear pump-to-Dirac-mass relation. A next step is to optimize the pump-dressed addressing fields jointly with the calibrated emitter response.

Loss determines what a successful pulse means in practice. At the passive exceptional point, complete conditional conversion can accompany the loss of most excitations, so a shorter conversion time alone does not establish better performance. Controlled nonunitary experiments and Liouvillian dynamics show why the measured ensemble and generator must be specified.\cite{CNNHXiao2017,CNNHWu2019,CNNHSong2019,CNNHGao2025} Exceptional degeneracies in coupled-mode and non-Bloch settings also refer to particular operators.\cite{CNNHDing2016,CNNHXiao2021} For bulk topology, biorthogonal geometry requires isolated spectral projectors,\cite{GarrisonWright1988,Esaki2011} and nonreciprocal models may need non-Bloch invariants and explicit boundary analysis.\cite{Lee2016,YaoWang2018,Kunst2018,Yokomizo2019,Okuma2020} Non-Hermitian Chern bands, spectral winding and geometry-dependent skin accumulation illustrate these issues.\cite{CNNHYao2018,CNNHZhang2020,CNNHZhang2022,CNNHXiao2020} The common-loss benchmark here preserves the Hermitian eigenvectors; our two-state transfer results do not establish a general non-Hermitian bulk--boundary correspondence.

Independent bond constructions, propagators, curvature integrals and density-matrix solvers reproduce the effective-model results. Lean~4 also checks 18 exact reduced-algebra statements: the complex RMS bound and its saturation, the matrix-square identity, self-energy reconstruction and passive-rate conditions among them. The supplied checks separate these algebraic guarantees from floating-point and FEM convergence evidence (Supplementary Sec.~S12). They make the control prediction and its assumptions reproducible. Material dispersion, RF electrodes, the selected atomic transitions and the full pumped-device mapping still require investigation.

\section{Conclusion}
For the specified Kagome Dirac interface, we find a sublattice-selective control principle. Retaining the interface spinors reveals why uniform Raman control cancels and selects a $1:1:-2$ pattern that saturates the diagonal-control RMS bound. Complete lattice calculations give target fidelity $0.999937$ at the tested smooth interface and resolve valley and leakage errors that spin-population measurements can miss. Passive dynamics gives the survival cost of conversion. An exploratory LN supercell reproduces the complex addressing pattern to $0.373\%$ and supplies nonlocal linear and EO response matrices. These results connect a reproducible projected control calculation to electromagnetic design targets. Realizing a complete device requires combining these targets with a calibrated spin-dependent interaction and loss model.

\section*{Supplementary Material}
The Supplementary Material contains the microscopic conventions, derivations, full-lattice construction, extended figures, convergence data, FEM assumptions and details of the Lean verification.
\section*{Author declarations and data availability}

\onecolumngrid
\clearpage
\twocolumngrid
\renewcommand{\bibsection}{\section*{References}}
\label{page:references}
\bibliography{references}
\end{document}